\documentclass[journal, twocolumn]{IEEEtran} 

\usepackage{amsmath}
\usepackage{amssymb}
\usepackage{graphics}
\usepackage{graphicx}
\usepackage{epsfig}
\usepackage[font=footnotesize,labelsep=period]{caption}
\usepackage[font=footnotesize,labelformat=simple]{subcaption}

\usepackage{color}
\usepackage{soul}
\usepackage{cite}
\usepackage{url}
\usepackage{cases}

\newcommand{\be}{\begin{equation}}

\newcommand{\ee}{\end{equation}}
\newcommand{\bea}{\begin{eqnarray}}
\newcommand{\eea}{\end{eqnarray}}
\newcommand{\bdp}{\begin{displaymath}}
\newcommand{\edp}{\end{displaymath}}

\newcommand{\hthickline}{\hlinewd{.8pt}}

\newcommand{\RN}[1]{%
  \textup{\uppercase\expandafter{\romannumeral#1}}%
}
\newcommand{\sss}{\scriptscriptstyle}

\makeatletter
\def\hlinewd#1{%
\noalign{\ifnum0=`}\fi\hrule \@height #1 \futurelet \reserved@a\@xhline}
\makeatother
\def\BibTeX{{\rm B\kern-.05em{\sc i\kern-.025em b}\kern-.08em
    T\kern-.1667em\lower.7ex\hbox{E}\kern-.125emX}}

\begin{document}
\title{\huge{Online Reinforcement Learning of X-Haul Content Delivery Mode in Fog Radio Access Networks}}

\author{\IEEEauthorblockN{\normalsize{Jihwan Moon, \textit{Member}, \textit{IEEE}, Osvaldo Simeone, \textit{Fellow}, \textit{IEEE}, Seok-Hwan Park, \textit{Member}, \textit{IEEE}, \\ and Inkyu Lee, \textit{Fellow}, \textit{IEEE}} \\
\small}
\thanks{This work was supported by the National Research Foundation through the Ministry of Science, ICT, and Future Planning (MSIP), Korean Government under Grant 2017R1A2B3012316. O. Simeone has received funding from the European Research Council (ERC) under the European Union's Horizon 2020 Research and Innovation Programme (Grant Agreement No. 725731). J. Moon and I. Lee are with the School of Electrical Engineering, Korea University, Seoul 02841, South Korea (e-mail: \{$\text{anschino}$, $\text{inkyu}$\}@korea.ac.kr).
O. Simeone is with the Department of Informatics, King's College London, London WC2R 2LS, U.K. (e-mail: $\text{osvaldo.simeone}$@kcl.ac.uk).
S.-H. Park is with the Division of Electronic Engineering, Chonbuk National
University, Jeonju 54896, South Korea (e-mail: $\text{seokhwan}$@jbnu.ac.kr).}
\vspace{-15pt}
}\maketitle \thispagestyle{empty}

\begin{abstract}
We consider a Fog Radio Access Network (F-RAN) with a Base Band Unit (BBU) in the cloud and multiple cache-enabled enhanced Remote Radio Heads (eRRHs). The system aims at delivering contents on demand with minimal average latency from a time-varying library of popular contents. Information about uncached requested files can be transferred from the cloud to the eRRHs by following either backhaul or fronthaul modes. The backhaul mode transfers fractions of the requested files, while the fronthaul mode transmits quantized baseband samples as in Cloud-RAN (C-RAN). The backhaul mode allows the caches of the eRRHs to be updated, which may lower future delivery latencies. In contrast, the fronthaul mode enables cooperative C-RAN transmissions that may reduce the current delivery latency. Taking into account the trade-off between current and future delivery performance, this paper proposes an adaptive selection method between the two delivery modes to minimize the long-term delivery latency. Assuming an unknown and time-varying popularity model, the method is based on model-free Reinforcement Learning (RL). Numerical results confirm the effectiveness of the proposed RL scheme.
\end{abstract}


\section{Introduction} \label{sec:1_intro}
The architecture of the recently launched fifth generation (5G) mobile system can leverage cloud processing at Base Band Units (BBUs), as well as edge processing, including edge caching, at enhanced Remote Radio Heads (eRRHs) \cite{YJKu:17}.
In order to enable a flexible functional split in this architecture, referred to as Fog-Radio Access Network (F-RAN) \cite{YYShih:17}, the concept of $\textit{X-haul}$ has been introduced to integrate the traditionally distinct backhaul and fronthaul connectivity modes for the interface between the BBU and the eRRH into a unified framework \cite{ADLOliva:15, TPfeiffer:15, NJGomes:15}. The backhaul mode enables the transfer of data packets from the BBU in the cloud to the eRRHs. In contrast, the fronthaul mode allows the BBU to carry out joint baseband processing and deliver quantized baseband samples to the eRRHs as in Cloud-RAN (C-RAN) \cite{HRen:18, JKim:19a, JKim:19b}.

In this work, we study an adaptive selection of backhaul and fronthaul transfer modes with the aim of optimizing the performance of content delivery. The content delivery in F-RANs has been widely studied in recent years \cite{SHPark:16a, SHPark:16b, ASengupta:17, JZhang:19, SMAzimi:18, ASadeghi:18, SOSomuyiwa:18}. Most studies assume offline caching with a static popularity model. Under these assumptions, references \cite{SHPark:16a} and \cite{SHPark:16b} investigated the problem of instantaneous delivery latency minimization and minimum data rate maximization, respectively, while keeping the contents of the caches fixed. In contrast, in \cite{ASengupta:17} and \cite{JZhang:19}, information-theoretic performance bounds were provided on the optimal high Signal-to-Noise-Ratio (SNR) performance by considering also the optimization of uncoded caching strategies. An extension of this work that accounts for time-varying and possibly unknown file popularity with online caching was described in \cite{SMAzimi:18}. Under an unknown dynamic popularity model, the works \cite{ASadeghi:18} and \cite{SOSomuyiwa:18} presented a Reinforcement Learning (RL) based optimization of online caching by assuming a backhaul mode.

\begin{figure}
\centering
\includegraphics[width=0.40\textwidth]{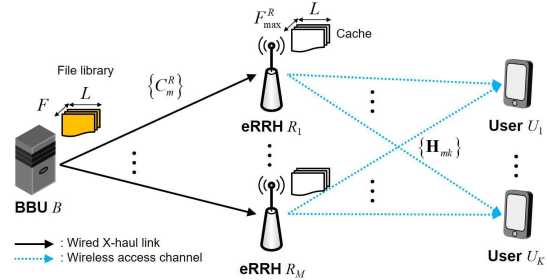}
\vspace{- 5 pt}
\caption{Illustration of the F-RAN system under study}
\label{fig:System_model}
\vspace{- 18 pt}
\end{figure}

In this paper, we investigate for the first time the online minimization of the long-term delivery latency over X-haul links in an F-RAN with time-varying unknown file popularity. We focus on the joint optimization of linear precoding strategies and the choice between fronthaul and backhaul modes. The backhaul mode enables cache updates at the eRRHs, hence potentially reducing future latencies. In contrast, the fronthaul mode allows cooperative C-RAN transmissions which decrease the current delivery latency \cite{SHPark:16a, SHPark:16b, ASengupta:17}. We propose a new model-free RL approach based on a linear value function approximation with properly selected features, and numerical results confirm the effectiveness of the proposed RL scheme.

$\textit{Notations}$: $\mathbb{E} \left[ \cdot \right]$ and $\text{Pr} \left( \cdot \right)$ stand for expectation and probability, respectively. $\left| \mathcal{A} \right|$ represents the cardinality of set $\mathcal{A}$, and $\mathbb{C}^{m \times n}$ denotes an $m \times n$ complex matrix. $\mathbb{I} \left\{ c \right\}$ outputs one if condition $c$ is true and zero otherwise. For a matrix $\textbf{X}$, $\left| \textbf{X} \right|$, $\textbf{X}^{T}$, $\textbf{X}^{H}$, $\textbf{X}^{- 1}$ and $\text{tr} \left( \textbf{X} \right)$ are defined as determinant, transpose, Hermitian, inverse and trace, respectively. $\textbf{I}_{\sss{m}}$ means an $m \times m$ identity matrix while $\otimes$ equals a Kronecker product operation. Also, $\text{diag} \big( \textbf{X}_{\sss{1}}, ..., \textbf{X}_{\sss{N}} \big)$ represents block-wise diagonalization of matrices $\textbf{X}_{\sss{1}}, ..., \textbf{X}_{\sss{N}}$. Lastly, $\mathcal{CN} \left( \boldsymbol{\mu}, \boldsymbol{\Omega} \right)$ indicates a circularly symmetric complex Gaussian distribution with mean vector $\boldsymbol{\mu}$ and covariance matrix $\boldsymbol{\Omega}$.

\section{System Model} \label{sec:2_system_model}
We study the F-RAN system illustrated in Fig. \ref{fig:System_model}, which consists of a BBU in the cloud, connected to $M$ cache-enabled eRRHs and $K$ users. Each X-haul link between the BBU and the $m$-th eRRH has capacity $C_{\sss{m}}^{R}$ bits per symbols and can be operated in both backhaul and fronthaul modes \cite{TPfeiffer:15}\cite{NJGomes:15}. The $k$-th user and the $m$-th eRRH are equipped with $N_{\sss{k}}^{U}$ and $N_{\sss{m}}^{R}$ antennas, respectively. We assume a time-slotted operation \cite{SOSomuyiwa:18}, and the wireless channel matrix $\textbf{H}_{\sss{mk}}$ between the $m$-th eRRH and the $k$-th user is assumed to be fixed for the given time scale of interest $T_{\sss{\text{B}}}$ slots. We also define $\mathcal{F} \triangleq \left\{ f_{\sss{1}}, ..., f_{\sss{F}} \right\}$ as the library of $F$ $L$-bit files, which may be requested by the users. Finally, we denote $\mathcal{F}^{R} ( t ) \subseteq \mathcal{F}$ as the subset of files cached at time slot $t$ at the eRRHs whose cardinality is bounded by $F_{\sss{\max}}^{R}$ files due to storage capacity constraints. Note that in this letter, we make a simplifying assumption that all the eRRHs store the same files in their respective caches. Generalization of the framework is possible but at the cost of a more cumbersome notation. Detailed request, online caching and delivery models are described in the following.

\subsection{Request Model and Online Caching} \label{sec:2.1_system_model_request}
In each time slot $t$, a subset $\mathcal{F}_{\sss{\text{pop}}} ( t ) \subseteq \mathcal{F}$ of files is popular in the sense that all users request files from $\mathcal{F}_{\sss{\text{pop}}} ( t )$. Specifically, the $k$-th user requests a uniformly selected file $f_{\sss{k}}^{U} ( t )$ from subset $\mathcal{F}_{\sss{\text{pop}}} ( t )$ without replacement \cite{SMAzimi:18}. The assumption of no replacement ensures that all requested files are distinct, yielding a worst-case performance analysis \cite{ASengupta:17}. We assume that the popularity $\mathcal{F}_{\sss{\text{pop}}} ( t )$ varies as a Markov process as in \cite{ASadeghi:18, RPedarsani:16, NGarg:19, ASadeghi:19}. This is a standard assumption which provides a first-order approximation of the evolution of the content popularity \cite{MZorziy:95}\cite{GCarofiglio:11}.
Let $\mathcal{K}_{\sss{\text{req}, \text{C}}} ( t )$ and $\mathcal{K}_{\sss{\text{req}, \text{NC}}} ( t )$ denote the indices of the users whose requested files $\mathcal{F}_{\sss{\text{req}, \text{C}}} ( t ) \triangleq \left\{ f_{\sss{k}}^{U} ( t ) \right\}_{\sss{k \in \mathcal{K}_{\sss{\text{req}, \text{C}}} ( t )}}$ are cached and the indices of users whose requested files $\mathcal{F}_{\sss{\text{req}, \text{NC}}} ( t ) \triangleq \left\{ f_{\sss{k}}^{U} ( t ) \right\}_{\sss{k \in \mathcal{K}_{\sss{\text{req}, \text{NC}}} ( t )}}$ are not cached at time $t$, respectively. In case the backhaul mode is selected at time slot $t$, the requested but uncached files in $\mathcal{F}_{\sss{\text{req}, \text{NC}}} ( t )$ are sent on all the X-haul links and cached. In order to make space for a new file, a previously cached file is evicted by following the standard Least Recently Used (LRU) rule \cite{TJohnson:94}.

\subsection{Delivery Operation} \label{sec:2.2_system_model_operation}
At each slot $t$, the X-haul link is used in either fronthaul or backhaul mode for $\Delta^{R} ( t, \text{a} ( t ) )$ symbols, where $\text{a} ( t ) = 0$ and $1$ indicate the selection of fronthaul and backhaul modes, respectively. Subsequently, the eRRHs deliver the requested files in set $\mathcal{F}_{\sss{\text{req}}} ( t ) \triangleq \mathcal{F}_{\sss{\text{req}, \text{C}}} ( t ) \cup \mathcal{F}_{\sss{\text{req}, \text{NC}}} ( t )$ over the wireless channel for $\Delta^{U} ( t, \text{a} ( t ) )$ symbols, based on the signals received on the X-haul links and on the cached contents. This results in a total latency of $\Delta ( t, \text{a} ( t ) ) = \Delta^{R} ( t, \text{a} ( t ) ) + \Delta^{U} ( t, \text{a} ( t ) )$ symbols for time slot $t$. Note that the eRRHs' caches are updated according to the caching mechanism described in Section~\ref{sec:2.1_system_model_request} only if the backhaul mode is selected as $\text{a} ( t ) = 1$.

\subsection{Problem Formulation} \label{sec:2.3_system_model_problem}
The delivery time $\Delta ( t, \text{a} ( t ) )$ at slot $t$ depends on the state of the system $\text{s} ( t ) = \{ \mathcal{F}_{\sss{\text{pop}}} ( t )$, $\mathcal{F}^{R} ( t )$, $\mathcal{F}_{\sss{\text{req}}} ( t ) \}$, which includes the set of popular files, cached files and requested files, respectively. Given the Markovity of the process $\mathcal{F}_{\text{pop}} ( t )$, the state $\text{s} ( t )$ evolves as a controlled Markov process. $\text{s} ( t )$ is partially observable since the set $\mathcal{F}_{\sss{\text{pop}}} ( t )$ is unknown, and it is only observed indirectly via the file set $\mathcal{F}_{\sss{\text{req}}} ( t )$. In particular, at time $t$, only the history of observations $\text{o} ( 1 \text{:} t ) \triangleq \left\{ \text{o} ( 1 ), ..., \text{o} ( t ) \right\}$ with $\text{o} ( t ) = \{ \mathcal{F}_{\sss{\text{req}}} ( t )$, $\mathcal{F}^{R} ( t ) \}$ is available to the system. Thus, a general policy can map the observations $\text{o} ( 1 \text{:} t )$ to the selected action $\text{a} ( t )$ through a conditional distribution $\pi ( \text{a} ( t ) | \text{o} ( 1 \text{:} t ) )$.

In this work, we aim at minimizing the average long-term delivery latency of the proposed F-RAN system over the selection of policy $\pi ( \text{a} ( t ) | \text{o} ( 1 \text{:} t ) )$. Given a forgetting factor $\gamma \leq 1$, the problem can be formulated as
\begin{subequations}
\begin{align}
	\text{(P):}  \min\limits_{\pi}  \mathbb{E}_{\sss{\pi}} \left[ \sum\nolimits_{t = 1}^{\infty}\gamma^{t}\Delta ( t, \text{a} ( t ) ) \right] 
	\text{s.t.} ~ \text{a} ( t ) \in \left\{ 0, 1 \right\}, \forall t,
\end{align}
\end{subequations}
where calculation of the total latency $\Delta ( t, \text{a} ( t ) )$ will be reviewed in Section~\ref{sec:3_instant}. The expectation in (P) is over the state distribution, which depends on the policy.

\section{Minimum Instantaneous Latency} \label{sec:3_instant}
In this section, we discuss how to evaluate the delivery latency $\Delta ( t, \text{a} ( t ) )$ in problem (P). We emphasize that $\Delta ( t, \text{a} ( t ) )$ for $\text{a} ( t ) = 0$ and $1$ is assumed known when solving problem (P) at each time slot $t$, and is derived as defined in this section. Following \cite{SHPark:16a}, we omit the time index $t$ for simplicity.

\subsection{Backhaul Mode} \label{sec:3.1_instant_backahul}
In the backhaul mode ($\text{a} = 1$), the BBU first fetches the requested but uncached files $\mathcal{F}_{\sss{\text{req}, \text{NC}}}$ and transmits them to the eRRHs. The backhaul transmission to the $m$-th eRRH takes $\Delta_{\sss{m}}^{R} = \big| \mathcal{F}_{\sss{\text{req}, \text{NC}}} \big| L / C_{\sss{m}}^{R}$ symbols, and the total backhaul latency is $\Delta^{R} = \max\nolimits_{m}\Delta_{\sss{m}}^{R}$, since all the eRRHs need to receive the files in $\mathcal{F}_{\sss{\text{req}, \text{NC}}}$. As a result, all the requested files in $\mathcal{F}_{\sss{\text{req}}}$ are available at the eRRHs and cooperative transmission across all eRRHs is feasible. Each file $f_{\sss{k}}^{U} \in \mathcal{F}_{\sss{\text{req}}}$ for the $k$-th user is encoded by each eRRH as the signal $\textbf{s}_{\sss{k}} \in \mathbb{C}^{n_{\sss{k}} \times 1} \sim \mathcal{CN} \left( \textbf{0}, \textbf{I}_{\sss{n_{\sss{k}}}} \right)$, where $n_{\sss{k}} \leq N_{\sss{k}}^{U}$ denotes the number of data streams allocated to the $k$-th user, which is assumed to be a fixed parameter. The transmit signal from the $m$-th eRRH is then given as $\textbf{x}_{\sss{m}} = \sum\nolimits_{k \in \mathcal{K}_{\sss{\text{req}}}}\textbf{G}_{\sss{mk}}\textbf{s}_{\sss{k}}$ where $\mathcal{K}_{\sss{\text{req}}} \triangleq \mathcal{K}_{\sss{\text{req}, \text{C}}} \cup \mathcal{K}_{\sss{\text{req}, \text{NC}}}$, and $\textbf{G}_{\sss{mk}} \in \mathbb{C}^{N_{\sss{m}}^{R} \times n_{\sss{k}}}$ is the precoding matrix for $\textbf{s}_{\sss{k}}$ at the $m$-th eRRH. Accordingly, the achievable rate for the $k$-th user on the wireless channel can be written as \cite{SHPark:16a}
\begin{align}
	R_{\sss{\text{back}, k}}^{U} \left( \left\{ \textbf{G}_{\sss{k}} \right\} \right) = \log_{2} \big| \textbf{I}_{\sss{N_{k}^{U}}} + \boldsymbol{\Phi}_{\sss{\text{back}, k}}^{U} \big| \ \text{[bits/symbol]},
\end{align}
where we have $\boldsymbol{\Phi}_{\sss{\text{back}, k}}^{U} \triangleq \big( \sum\nolimits_{\ell \in \mathcal{K}_{\sss{\text{req}}} \backslash k}\textbf{H}_{\sss{k}}\textbf{G}_{\sss{\ell}}\textbf{G}_{\sss{\ell}}^{H}\textbf{H}_{\sss{k}}^{H} + \sigma_{\sss{k}}^{2}\textbf{I}_{\sss{N_{k}^{U}}} \big)^{- 1} \textbf{H}_{\sss{k}}\textbf{G}_{\sss{k}}\textbf{G}_{\sss{k}}^{H}\textbf{H}_{\sss{k}}^{H}$ with $\textbf{H}_{\sss{k}} \triangleq \big[ \textbf{H}_{\sss{1k}} \cdots \textbf{H}_{\sss{Mk}} \big]$ and $\textbf{G}_{\sss{k}} \triangleq \big[ \textbf{G}_{\sss{1k}}^{T} \cdots \textbf{G}_{\sss{Mk}}^{T} \big]^{T}$, and $\sigma_{\sss{k}}^{2}$ represents the additive white Gaussian noise variance at the $k$-th user.

The latency $\Delta_{\sss{k}}^{U}$ for delivering file $f_{\sss{k}}^{U}$ for the $k$-th user is obtained as $\Delta_{\sss{k}}^{U} = L / R_{\sss{\text{back}, k}}^{U} \left( \left\{ \textbf{G}_{\sss{k}} \right\} \right)$, and the overall wireless channel latency equals $\Delta^{U} = \max\nolimits_{k}\Delta_{\sss{k}}^{U}$, since every requesting user needs to receive the requested file. The minimum instantaneous latency $\Delta$ for $\text{a} = 1$ can hence be found as a solution of the problem
\begin{subequations}
\begin{align}
	\text{(P1):} ~ &\min\nolimits_{\Delta^{U}, \left\{ \textbf{G}_{\sss{k}} \right\}} ~ \Delta^{R} + \Delta^{U} \\[- 2 pt]
	\text{s.t.} ~~~ &\Delta^{U} \geq {L} / {R_{\sss{\text{back}, k}}^{U} \left( \left\{ \textbf{G}_{\sss{k}} \right\} \right)}, ~ \forall k \in \mathcal{K}_{\sss{\text{req}}}, \\[- 2 pt]
	&\text{tr} \big( \sum\nolimits_{k \in \mathcal{K}_{\sss{\text{req}}}} \!\!\!\! \textbf{E}_{\sss{m}}\textbf{G}_{\sss{k}}\textbf{G}_{\sss{k}}^{H}\textbf{E}_{\sss{m}}^{H} \big) \leq P_{\sss{m}}^{R}, m = 1, ..., M,
\end{align}
\end{subequations}
where $P_{\sss{m}}^{R}$ denotes the maximum transmit power of the $m$-th eRRH, and we define $\textbf{E}_{\sss{m}} \triangleq \big[ \textbf{0} \cdots \textbf{I}_{\sss{N_{m}^{R}}} \cdots \textbf{0} \big]$ in which an identity matrix $\textbf{I}_{\sss{N_{m}^{R}}}$ spans columns from $\sum\nolimits_{\ell = 1}^{m - 1}N_{\sss{\ell}}^{R} + 1$ to $\sum\nolimits_{\ell = 1}^{m}N_{\sss{\ell}}^{R}$. Although problem (P1) is jointly non-convex, a stationary point can be attained by leveraging Successive Convex Approximation (SCA) as detailed in \cite{SHPark:16a}.

\subsection{Fronthaul Mode} \label{sec:3.2_instant_fronthaul}
Under the fronthaul mode, any requested but uncached file $f_{\sss{k}}^{U} \in \mathcal{F}_{\sss{\text{req}, \text{NC}}}$ for the $k$-th user is jointly encoded and precoded at the BBU. The resulting signal dedicated for the $m$-th eRRH is written as $\hat{\textbf{x}}_{\sss{m}} = \sum\nolimits_{k \in \mathcal{K}_{\sss{\text{req}, \text{NC}}}} \textbf{W}_{\sss{mk}}\textbf{s}_{\sss{k}}$, where $\textbf{s}_{\sss{k}} \in \mathbb{C}^{n_{\sss{k}} \times 1} \sim \mathcal{CN} \left( \textbf{0}, \textbf{I}_{\sss{n_{\sss{k}}}} \right)$ encodes file $f_{\sss{k}}^{U}$, and $\textbf{W}_{\sss{mk}} \in \mathbb{C}^{N_{\sss{m}}^{R} \times n_{\sss{k}}}$ represents the corresponding precoding matrix for the $m$-th eRRH. The BBU then performs compression on $\hat{\textbf{x}}_{\sss{m}}$ prior to transferring to the eRRHs. As a result, the decompressed signal at the $m$-th eRRH can be written by $\tilde{\textbf{x}}_{\sss{m}} = \hat{\textbf{x}}_{\sss{m}} + \textbf{q}_{\sss{m}}$ with quantization noise $\textbf{q}_{\sss{m}} \in \mathbb{C}^{N_{\sss{m}}^{R} \times 1} \in \mathcal{CN} \left( \textbf{0}, \boldsymbol{\Omega}_{\sss{m}} \right)$ for a given covariance matrix $\boldsymbol{\Omega}_{\sss{m}}$ \cite{SHPark:16a}\cite{SHPark:16b}.

The rest of the requested cached files $\mathcal{F}_{\sss{\text{req}, \text{C}}}$ are locally precoded with $\left\{ \textbf{G}_{\sss{mk}} \right\}$ at the eRRHs in the same manner as in the backhaul mode. The final transmit signal at the $m$-th eRRH is then given as $\textbf{x}_{\sss{m}} = \sum\nolimits_{k \in \mathcal{K}_{\sss{\text{req}, \text{C}}}} \textbf{G}_{\sss{mk}}\textbf{s}_{\sss{k}} + \tilde{\textbf{x}}_{\sss{m}}$, and the achievable rate for the $k$-th user can be obtained as \cite{SHPark:16a}
\begin{align}
	R_{\sss{\text{front}, k}}^{U} \big( \big\{ \tilde{\textbf{G}}_{\sss{k}} \big\}, \boldsymbol{\Omega}_{\sss{R}} \big) = \log_{2} \big| \textbf{I}_{\sss{N_{k}^{U}}} + \boldsymbol{\Phi}_{\sss{\text{front}, k}}^{U} \big| \ \text{[bits/symbol]},
\end{align}
where we have $\boldsymbol{\Phi}_{\sss{\text{front}, k}}^{U} \triangleq \big( \sum\nolimits_{\ell \in \mathcal{K}_{\sss{\text{req}}} \backslash k}\textbf{H}_{\sss{k}}\tilde{\textbf{G}}_{\sss{\ell}}\tilde{\textbf{G}}_{\sss{\ell}}^{H}\textbf{H}_{\sss{k}}^{H} + \textbf{H}_{\sss{k}}\boldsymbol{\Omega}_{\sss{R}}\textbf{H}_{\sss{k}}^{H} + \sigma_{\sss{k}}^{2}\textbf{I}_{\sss{N_{k}^{U}}} \big)^{- 1} \textbf{H}_{\sss{k}}\tilde{\textbf{G}}_{\sss{k}}\tilde{\textbf{G}}_{\sss{k}}^{H}\textbf{H}_{\sss{k}}^{H}$, $\boldsymbol{\Omega}_{\sss{R}} \triangleq \text{diag} \big( \boldsymbol{\Omega}_{\sss{1}}, ..., \boldsymbol{\Omega}_{\sss{M}} \big)$, $\tilde{\textbf{G}}_{\sss{k}} \triangleq \big[ \tilde{\textbf{G}}_{\sss{1k}}^{T} \cdots \tilde{\textbf{G}}_{\sss{Mk}}^{T} \big]^{T}$ with $\tilde{\textbf{G}}_{\sss{mk}} \triangleq b_{\sss{k}}^{U}\textbf{G}_{\sss{mk}} + \left( 1 - b_{\sss{k}}^{U} \right) \textbf{W}_{\sss{mk}}$, and $b_{\sss{k}}^{U} = 1$ if $f_{\sss{k}}^{U} \in \mathcal{K}_{\sss{\text{req}, \text{C}}}$ and $b_{\sss{k}}^{U} = 0$ otherwise for the $k$-th user.

The wireless channel latency $\Delta^{U}$ is defined in the same way as in the backhaul mode. For the fronthaul latency, by the rate-distortion theory, sending quantized signals to the $m$-th eRRH consumes
\begin{align}
	g_{\sss{m}} \big( \big\{ \tilde{\textbf{G}}_{\sss{k}} \big\}, \boldsymbol{\Omega}_{\sss{R}} \big) = \log_{2} \big| \textbf{I}_{\sss{N_{m}^{R}}} + \boldsymbol{\Phi}_{\sss{m}}^{R} \big| \ \text{[bits/symbol]},
\end{align}
with $\boldsymbol{\Phi}_{\sss{m}}^{R} \triangleq \big( \textbf{E}_{\sss{m}} \boldsymbol{\Omega}_{\sss{R}}\textbf{E}_{\sss{m}}^{H} \big)^{- 1}\sum\nolimits_{k \in \mathcal{K}_{\sss{\text{req}, \text{NC}}}}\textbf{E}_{\sss{m}}\tilde{\textbf{G}}_{\sss{k}}\tilde{\textbf{G}}_{\sss{k}}^{H}\textbf{E}_{\sss{m}}^{H}$ \cite{SHPark:16a}. Compressing $\Delta^{U}$ symbols produces $\Delta^{U} g_{\sss{m}} \big( \big\{ \tilde{\textbf{G}}_{\sss{k}} \big\}, \boldsymbol{\Omega}_{\sss{R}} \big)$ bits, which need to be transferred from the BBU to the $m$-th eRRH. Therefore, the fronthaul latency is given by $\Delta^{R} = \max\nolimits_{m}\Delta_{\sss{m}}^{R}$ where $\Delta_{\sss{m}}^{R} = \Delta^{U} g_{\sss{m}} \big( \big\{ \tilde{\textbf{G}}_{\sss{k}} \big\}, \boldsymbol{\Omega}_{\sss{R}} \big) / C_{\sss{m}}^{R}$, and the minimum instantaneous latency $\Delta$ for $\text{a} = 0$ is calculated as a solution of the problem
\begin{subequations}
	\begin{align}
	\text{(P2):} ~ &\min\nolimits_{\Delta^{R}, \Delta^{U}, \left\{ \tilde{\textbf{G}}_{\sss{k}} \right\}, \boldsymbol{\Omega}_{\sss{R}}} ~ \Delta^{R} + \Delta^{U} \\[- 0 pt]
	\text{s.t.} ~~~ &\Delta^{R} \geq {\Delta^{U} g_{\sss{m}} \big( \big\{ \tilde{\textbf{G}}_{\sss{k}} \big\}, \boldsymbol{\Omega}_{\sss{R}} \big)} / {C_{\sss{m}}^{R}}, m = 1, ..., M, \\[- 0 pt]
	&\Delta^{U} \geq {L} / {R_{\sss{\text{front}, k}}^{U} \big( \big\{ \tilde{\textbf{G}}_{\sss{k}} \big\}, \boldsymbol{\Omega}_{\sss{R}} \big)}, ~ \forall k \in \mathcal{K}_{\sss{\text{req}}}, \\[- 0 pt]
	&\text{tr} \big( \sum\nolimits_{k \in \mathcal{K}_{\sss{\text{req}}}}\textbf{E}_{\sss{m}}\tilde{\textbf{G}}_{\sss{k}}\tilde{\textbf{G}}_{\sss{k}}^{H}\textbf{E}_{\sss{m}}^{H} + \textbf{E}_{\sss{m}}\boldsymbol{\Omega}_{\sss{R}}\textbf{E}_{\sss{m}}^{H} \big) \leq P_{\sss{m}}^{R}, \nonumber\\[- 0 pt]
	& ~ m = 1, ..., M,
	\end{align}
\end{subequations}
which can be tackled via the SCA approach detailed in \cite{SHPark:16a}. The total worst-case order of complexity for the SCA method can be expressed as $\mathcal{O} ( N_{\sss{\text{SCA}}} \sqrt{N_{\sss{\text{const}}}}\log ( N_{\sss{\text{const}}} / \epsilon ) )$ where $\epsilon$, $N_{\sss{\text{SCA}}}$ and $N_{\sss{\text{const}}}$ indicate the desired error tolerance, the maximum number of the SCA iterations and the number of constraints, respectively \cite{SBoyd:04}. Here, $N_{\sss{\text{const}}}$ equals $\left| \mathcal{K}_{\sss{\text{req}}} \right| + M$ in (P1) and $\left| \mathcal{K}_{\sss{\text{req}}} \right| + 2M$ in (P2).

\section{RL-Based X-Haul Online Optimization} \label{sec:4_RL}
In this section, we solve problem (P) by proposing an online on-policy RL-based optimization strategy \cite{RSSutton:18}.

\subsection{Problem (P) as a Partially Observable Decision Process} \label{sec:4.1_RL_MDP}
As discussed in Section~\ref{sec:2_system_model}, problem (P) is a Partially Observable Markov Decision Process (POMDP) with the action space $\left\{ 0, 1 \right\}$ and the instantaneous reward given by the negative latency $\text{r} ( t + 1 ) = - \Delta ( t, \text{a} ( t ) )$. In order to reduce the complexity of the policy, we optimize here over memoryless policies that select an action $\text{a} ( t )$ based only on the latest observation $\text{o} ( t )$ at time slot $t$ \cite{MLLittman:94}\cite{YLi:11} as well as a summary of the previous observations $\text{o} ( 1 \text{:} t )$ given by the set $\{ \tau_{\sss{\text{req}, f}} ( t ) \}_{\sss{f \in \mathcal{F}^{R} ( t )}} \}$ where $\tau_{\sss{\text{req}, f}} ( t )$ is the most recent time slot at which cached file $f$ was requested at time slot $t$.

\subsection{SARSA with Linear Value Function Approximation} \label{sec:4.2_RL_linear}
To optimize over memoryless policies, we adopt the online on-policy value-based strategy State-Action-Reward-State-Action (SARSA) with a carefully designed linear approximation \cite{RSSutton:18}. The SARSA updates an action-value function, or Q-function, $q \left( o, a \right)$ that estimates the expected return $\mathbb{E} [ G ( t ) | \text{o} = o, \text{a} = a ]$ with $G ( t ) \triangleq \sum\nolimits_{\tau = 0}^{\infty}\gamma^{\tau}\text{r} ( t + \tau + 1 )$. Since the total size of the observation space in (P) grows exponentially with $F$, we propose a linear value function approximation $\hat{q} \left( o, a, \textbf{w} \right) \triangleq \textbf{w}^{T}\textbf{x} \left( o, a \right)$, where $\textbf{w}$ is a parameter vector to be learned, and $\textbf{x} \left( o, a \right)$ denotes a feature vector representing the observation-action pair $\left( o, a \right)$ \cite{RSSutton:18}.

In order to determine a suitable feature vector, we first note that vector $\textbf{x} \left( o, a \right)$ should contain sufficient information to quantify the value of caching for currently cached and requested files. Frequently requested files typically yield lower future latencies when cached, but an optimal choice should account not only for their popularity but also for their remaining $\textit{life}$ $\textit{time}$, which is a duration that a file remains popular (see Sec.~II of \cite{STraverso:13} for further discussion).

Based on these considerations, we introduce a variable $\phi_{\sss{\ell}} ( t )$ for every file $f_{\sss{\ell}} \in \mathcal{F}$ as a function of the current observation $\text{o} ( t )$ at time slot $t$. We set it as $\phi_{\sss{\ell}} ( t ) = 1$ if $f_{\sss{\ell}} \in \mathcal{F}_{\sss{\text{req}, \text{NC}}} ( t )$, $\phi_{\sss{\ell}} ( t ) = 2$ if $f_{\sss{\ell}} \in \mathcal{F}^{R} ( t )$ and $\phi_{\sss{\ell}} ( t ) = 0$ otherwise. Furthermore, we also include a variable $\theta ( t ) \triangleq t - \max\nolimits_{f \in \mathcal{F}^{R} ( t )}\tau_{\sss{\text{req}, f}}$ that measures the ``age'' of the currently cached files, that is, the maximum time elapsed since the last request of the cached files. We can quantize this variable by $N_{\sss{\Theta}}$ ranges $\Theta_{\sss{1}}, ..., \Theta_{\sss{N_{\sss{\Theta}}}} \subseteq \mathbb{R}^{+}$ with $\Theta_{\sss{i}} \cap \Theta_{\sss{j}} = \emptyset$ for all $i \neq j$ and $\bigcup \Theta_{\sss{i}} = \mathbb{R}^{+}$. If the caches are up to date, the quantity $t - \tau_{\sss{\text{req}, f}}$ is small for all $f \in \mathcal{F}^{R} ( t )$, and hence $\theta ( t )$ is also small. Otherwise, if there exists any file $f \in \mathcal{F}^{R} ( t )$ with large $t - \tau_{\sss{\text{req}, f}}$, a refresh of the caches may be required.

Using the variables introduced above, we define the feature vector $\textbf{x} \left( o ( t ), a ( t ) \right)$ as
\begin{align}
	\label{eq:x_feature}
	\textbf{x} \left( o ( t ), a ( t ) \right) = \big[ \boldsymbol{\phi}_{\sss{1}}^{T} ( t ) \ \cdots \ \boldsymbol{\phi}_{\sss{F}}^{T} ( t ) \ \boldsymbol{\theta}^{T} ( t ) \big]^{T} \otimes \textbf{a} ( t ),
\end{align}
where we have used the one-hot encoded vectors $\boldsymbol{\phi}_{\sss{\ell}} ( t ) \triangleq [ \mathbb{I} \{ \phi_{\sss{\ell}} ( t ) = 1 \} \ \mathbb{I} \{ \phi_{\sss{\ell}} ( t ) = 2 \} \ \mathbb{I} \{ \phi_{\sss{\ell}} ( t ) = 0 \} ]^{T}$, $\boldsymbol{\theta} ( t ) \triangleq [ \mathbb{I} \{ \theta ( t ) \in \Theta_{\sss{1}} \} \ \cdots \ \mathbb{I} \{ \theta ( t ) \in \Theta_{\sss{N_{\Theta}}} \} ]^{T}$ and $\textbf{a} ( t ) \triangleq [ \mathbb{I} \{ \text{a} ( t ) = 0 \} \ \mathbb{I} \{ \text{a} ( t ) = 1 \} ]^{T}$. The feature vector $\textbf{x} \left( o \left( t \right), a \left( t \right) \right)$ in (\ref{eq:x_feature}) has dimension $2 ( N_{\sss{\Theta}} + 3F )$, which increases linearly in $F$ and is hence significantly smaller than the size of the conventional look-up table-based SARSA. The effectiveness of the proposed feature vector $\textbf{x} \left( \text{o} ( t ), a \left( t \right) \right)$ will be verified in Section~\ref{sec:5_numerical_results}.

The overall proposed procedure for solving (P) is summarized in Algorithm~$1$ where $\delta ( t, \textbf{w} ) \triangleq \text{r} ( t + 1 ) + \gamma \hat{q} \left( \text{o} ( t + 1 ), \text{a} ( t + 1 ), \textbf{w} \right) - \hat{q} \left( o, a, \textbf{w} \right)$ denotes the temporal difference error, and $\textbf{E}$ indicates the eligibility trace. Here, an $\epsilon$-greedy exploration strategy with decreasing $\epsilon$ is adopted. Note that $\textbf{E}$ is used to assign credit for the current reward to the most frequently visited states and selected actions, so as to enable online learning (see \cite{RSSutton:18} for details).

\vspace{- 10 pt}

\begin{table}[ht]
	\small
	\begin{align}
	\begin{array}{l}
		\hthickline
		\text{Algorithm 1: Proposed RL-based solution for problem (P)} \\
		\hthickline
		\text{Initialize the total number of episodes $N_{\sss{\text{epi}}}$, weight vector $\textbf{w} = \textbf{0}$,} \\
		\text{eligibility trace $\textbf{E} = \textbf{0}$, and parameter $\gamma, \lambda \in \left( 0, 1 \right]$} \\
		\text{For $n_{\sss{\text{epi}}} = 1:N_{\sss{\text{epi}}}$} \\
		~~ \text{Randomly initialize cached contents $\mathcal{F}^{R} \left( 0 \right)$ and generate $\left\{ \textbf{H}_{\sss{mk}} \right\}$} \\
		~~ \text{For $t = 1:T_{\sss{\text{B}}}$} \\
		~~ ~~ \text{Collect observation $\text{o} \! \left( t \right) \!\! = \!\! \{ \! \mathcal{F}_{\sss{\text{req}}} ( t )$, $\mathcal{F}^{R} ( t )$,$\{ \tau_{\sss{\text{req}, f}} ( t ) \}_{\sss{f \in \mathcal{F}^{R} ( t )}} \! \}$} \\
		~~ ~~ \text{Choose the delivery mode greedily with probability $1 \!\! - \!\! 1/n_{\sss{\text{epi}}}$} \\
		~~ ~~ \text{as $\text{a} ( t ) = \arg\max\nolimits_{a'} \textbf{w}^{T}\textbf{x} \left( \text{o} ( t ), a' \right)$, and uniformly with} \\
		~~ ~~ \text{probability $1/n_{\sss{\text{epi}}}$} \\
		~~ ~~ \text{If $\text{a} ( t ) = 1$, update $\mathcal{F}_{\sss{\text{cache}}, R} \left( t \right)$ according to LRU} \\
		~~ ~~ \text{Set $\text{r} ( t + 1 ) = - \Delta ( t, \text{a} ( t ) )$} \\
		~~ ~~ \text{Update $\textbf{E} \leftarrow \gamma\lambda\textbf{E} + \textbf{x} \left( o, a \right)$} \\
		~~ ~~ \text{Update $\textbf{w} \leftarrow  \textbf{w} + \beta\delta ( t, \textbf{w} ) \textbf{E}$ with $\beta = 1 / n_{\sss{\text{epi}}}$} \\
		~~ \text{End} \\
		\text{End} \\
		\hthickline
	\end{array} \nonumber
	\end{align}
\end{table}

\vspace{- 20 pt}

\section{Numerical Results} \label{sec:5_numerical_results}
In this section, the performance of the proposed RL-based algorithm is evaluated via numerical examples. We adopt the channel model $\textbf{H}_{\sss{mk}} = \sqrt{\rho_{\sss{mk}}} \hat{\textbf{H}}_{\sss{mk}}$, where $\rho_{\sss{mk}} \triangleq \rho_{\sss{0}} \big( \frac{d_{\sss{mk}}}{d_{\sss{0}}} \big)^{- \eta}$ equals the distance-dependent path loss between eRRH $R_{\sss{m}}$ and user $U_{\sss{k}}$, $\rho_{\sss{0}}$ indicates the path loss at reference distance $d_{\sss{0}}$, $\eta$ is the path loss exponent, and $d_{\sss{mk}}$ represents the distance between the $m$-th eRRH and the $k$-th user. Each element of $\hat{\textbf{H}}_{\sss{mk}}$ follows an independent complex Gaussian distribution with zero mean and unit variance. The eRRHs and the users are circularly placed from the BBU at the center with uniformly distributed angles and distance $d_{\sss{BR}} = 200$~m and $d_{\sss{BU}} = 400$~m, respectively. The bandwidth is $20$~MHz and the thermal noise is $- 170$~dBm/Hz. We set $K = 10$, $M = 3$, $\rho_{\sss{0}} = 10^{- 3}$, $d_{\sss{0}} = 1$~m, $\eta = 3.5$, $T_{\sss{\text{B}}} = 100$ time slots, $F_{\sss{\max}}^{R} = 4$ files, $P_{\sss{m}}^{R} = 30$~dBm, $N_{\sss{m}}^{R} = N_{\sss{k}}^{U} = 1$ and $C_{\sss{m}}^{R} = 0.1$~bits per symbol. For RL, we use the hyperparameters $\gamma = 1$, $\lambda = 0.5$, and $\Theta_{\sss{\ell}} = [2(\ell - 1), \min ( 2(\ell - 1) + 1, \theta_{\sss{\max}} )]$ with $N_{\sss{\Theta}} = 11$ where $\theta_{\sss{\max}} = 20$ limits the maximum value of $\theta ( t )$.

Reference \cite{STraverso:13} demonstrated that the popularity of files often exhibits temporal locality in the sense that the content is frequently requested in a bursty fashion for a certain life time. Motivated by these findings, we model the evolution of the subset $\mathcal{F}_{\sss{\text{pop}}} ( t )$ of popular files such that a currently unpopular file $f$ has a probability of $\text{P}_{\sss{\text{pop}, f}}$ to become popular, and file $f$ remains popular for $T_{\sss{\text{life}, f}}$ time slots. We assume Zipf's distribution \cite{DMWPowers:98} for $\text{P}_{\sss{\text{pop}, f_{\ell}}} = \ell^{- \xi} / \sum\nolimits_{\nu = 1}^{F}\nu^{- \xi}$ with $\xi = 1$. The proposed RL scheme is compared with a greedy fronthaul/backhaul mode selection that minimizes the current delivery latency at each time slot as well as with an offline scheme that keeps the $F_{\sss{\max}}^{R}$ most popular files with the largest $\text{P}_{\sss{\text{pop}, f}}$ under the idealized assumption that this is known in prior.

\begin{figure}
\centering
\includegraphics[width=0.48\textwidth]{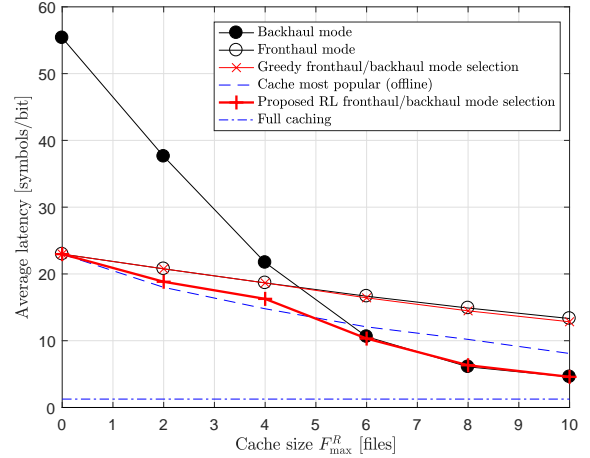}
\vspace{-5pt}
\caption{Average latency with respect to the maximum cache size $F_{\sss{\max}}^{R}$}
\label{fig:Plot_Main_avg_serial_F_R_bar}
\end{figure}

Fig.~\ref{fig:Plot_Main_avg_serial_F_R_bar} compares the average long-term latency performance as a function of the eRRHs' cache size $F_{\sss{\max}}^{R}$ for $P_{\sss{m}}^{R} = 30$~dBm, $T_{\sss{\text{life}, f}} = 10$ and $F = 20$. We also limit the maximum number of the SCA iterations for solving (P1) and (P2) as $N_{\sss{\text{SCA}}} = 10$. Note that the convergence to a stationary point for SCA does not affect the convergence of SARSA since we treat the negative reward function $- \Delta ( t, \text{a} ( t ) )$ as fixed. With $F_{\sss{\max}}^{R} \leq 4$, the fronthaul mode is seen to yield a lower latency than the backhaul mode given the limited advantage of caching in this regime. The opposite is true when the eRRHs have larger caches, such as $F_{\sss{\max}}^{R} > 4$, in which the backhaul mode outperforms the fronthaul mode. In agreement with the results in \cite{SHPark:16a, SHPark:16b, ASengupta:17} and \cite{SMAzimi:18}, the greedy scheme almost always selects the fronthaul mode and is hence strongly suboptimal for large enough $F_{\sss{\max}}^{R}$. The proposed RL method exhibits the lowest latency among all schemes that do not assume the knowledge of the popularity probability. It can be checked that the gain is not obtained by statically selecting the best mode at each time instant, but rather by carrying out an optimized dynamic selection. It is also observed that in a large $F_{\sss{\max}}^{R}$ regime, the proposed strategy can outperform the static offline scheme which assumes popularity dynamics to be known in advance.

\section{Conclusions} \label{sec:6_conclusion}
In this paper, we have demonstrated the advantage of adaptively selecting between the backhaul and fronthaul transfer modes as a function of the current cache contents and the history of past requests in an F-RAN system. The proposed RL-based strategy has been shown via numerical results to outperform baseline schemes, confirming the potential advantages of an X-haul implementation over static fronthaul or backhaul deployments.

\bibliographystyle{ieeetr}

\input{bibliography.filelist}

\end{document}